\documentclass[11pt,a4paper]{article}
\pdfoutput=1
\usepackage{jheppub}
\usepackage{mciteplus}
\usepackage{hyperref}
\hypersetup{colorlinks,urlcolor=blue,linktoc=page}
\usepackage{amssymb}
\usepackage{amsmath}
\usepackage{subfigure}
\usepackage{comment}
\usepackage{slashed}
\usepackage{bbm}
\usepackage{graphicx}
\usepackage{wrapfig}
\usepackage{caption}
\usepackage[UKenglish]{isodate}
\usepackage[english]{babel}
\usepackage[bottom]{footmisc}
\usepackage[normalem]{ulem}	

\usepackage{color}

\def\abs#1{\left|#1\right|}
\def\msol{m_\textup{sol}}
\def\matm{m_\textup{atm}}

\def\eref#1{(\ref{#1})}

\begin{document}

\subheader{\footnotesize\sc Preprint numbers: FTUAM-16-22, IFT-UAM/CSIC-16-055}
\title{A consistent model for leptogenesis, dark matter and the IceCube signal}

\author[a]{M.~Re~Fiorentin}
\author[b,c]{V.~Niro}
\author[d,e]{N.~Fornengo}

\affiliation[a]{School of Physics and Astronomy, University of Southampton, SO17 1BJ Southampton, U.K.
}
\affiliation[b]{Departamento de F\'isica Te\'orica, Universidad Aut\'onoma de Madrid, 
Cantoblanco, E-28049 Madrid, Spain
}
\affiliation[c]{
Instituto de F\'isica Te\'orica UAM/CSIC, Calle Nicol\'as Cabrera 13-15, Cantoblanco, E-28049 Madrid, Spain
}
\affiliation[d]{Dipartimento di Fisica, Universit\'a di Torino, via P. Giuria, 1, I-10125 Torino, Italy
}
\affiliation[e]{Istituto Nazionale di Fisica Nucleare, Sezione di Torino, via P. Giuria, 1, I-10125 Torino, Italy
}

\emailAdd{m.re-fiorentin@soton.ac.uk}
\emailAdd{viviana.niro@uam.es}
\emailAdd{nicolao.fornengo@unito.it}

\abstract{
We discuss a left-right symmetric extension of the Standard Model in which the three additional 
right-handed neutrinos play a central role in explaining the baryon asymmetry of the Universe, the dark matter 
abundance and the ultra energetic signal detected by the IceCube experiment. The energy spectrum and neutrino 
flux measured by IceCube are ascribed to the decays of the lightest right-handed neutrino $N_1$, thus fixing 
its mass and lifetime, while the production of  $N_1$ in the primordial thermal bath occurs via a freeze-in mechanism driven
by the additional $SU(2)_R$ interactions. The constraints imposed by IceCube and the dark matter 
abundance allow nonetheless the heavier right-handed neutrinos to realize a standard type-I seesaw leptogenesis, with the $B-L$ 
asymmetry dominantly produced by the next-to-lightest neutrino $N_2$. Further consequences and predictions of the model are that: the
$N_1$ production implies a specific power-law relation between the reheating temperature of the Universe and the 
vacuum expectation value of the $SU(2)_R$ triplet; leptogenesis imposes a lower bound on the reheating temperature
of the Universe at $7\times10^9\,\mbox{GeV}$. Additionally, the model requires a vanishing absolute neutrino mass scale $m_1\simeq0$.
}


\maketitle

\section{Introduction}

The IceCube experiment has so far reported evidence for extraterrestrial high-energy neutrinos~\cite{Aartsen:2013bka,icecube2,icecube1,Aartsen:2015rwa}, 
which cannot be explained by the atmospheric and prompt neutrino components~\cite{Garzelli:2015psa,Halzen:2016pwl,Halzen:2016thi}.  
It is possible to relate this signal to astrophysical sources, 
see for example Refs.~\cite{Murase:2014tsa,Anchordoqui:2013dnh,Cholis:2012kq} for a general discussion, and references therein\footnote{A lot 
of work has been done in trying to explain the IceCube events in term of astrophysical sources. We refer for example to the references contained 
in~\cite{Murase:2014tsa,Anchordoqui:2013dnh,Cholis:2012kq} for specific discussions.}, 
but it is certainly intriguing to speculate about a new-physics origin of these events. Some models have been proposed, explaining the high-energy IceCube signal as the decay 
of a long-lived particle \cite{moroi1,moroi2}, that can also constitute a viable dark matter (DM) 
candidate \cite{Feldstein:2013kka,Bhattacharya:2014vwa,Bhattacharya:2014yha,feldstein_yanagida,serpico1,serpico2,Bai:2013nga,murase,Fong:2014bsa,Kopp:2015bfa,
Esmaili:2015xpa,Anisimov:2008gg,kitano,extended_seesaw,Boucenna:2015tra, Chianese:2016opp,Barger:2013pla,Ko:2015nma,Dudas:2014bca,DiBari:2016guw,Dev:2016qbd}. 

A model of right handed neutrino DM predicting a signal in high energy neutrinos and able to reproduce the matter-antimatter asymmetry with leptogenesis was presented in Ref.~\cite{Anisimov:2008gg}. Ref.~\cite{kitano} applied a similar framework to interpret  the IceCube signal. These are also the aims of this work:
as in Ref.~\cite{Anisimov:2008gg,kitano}, we will make use of additional heavy right-handed (RH) neutrinos which are responsible for neutrino masses and mixing, 
via the seesaw mechanism~\cite{minkowski,yanagida_seesaw,gell-mann,glashow_seesaw,barbieri,mohapatra,PhysRevD.22.2227,PhysRevD.25.774}, and leptogenesis~\cite{fukugita_yanagida}. However, while in Ref.~\cite{kitano} the heavy neutrinos are produced by means of inflaton decay, here 
we will seek for a production mechanism directly from the thermal bath. 

We therefore consider a left-right symmetric model (LRSM)~\cite{lr_review1,lr_review2,senjanovic}, with gauge group $SU(2)_L\otimes SU(2)_R\otimes U(1)_{\tilde{Y}}$ 
in which three RH neutrinos $N_i$ are naturally accommodated into three RH doublets of $SU(2)_R$. 
Among the three heavy neutrinos, we choose the lightest one, $N_1$, to be quasi-stable, so that, if suitably long lived, it can be responsible for the IceCube signal through 
its decays into light neutrinos, as specified by the LRSM lagrangian. The energy spectrum of the IceCube events and the measured flux determine the mass and the lifetime of $N_1$.
We will show that hadronic decays of $N_1$ induce an additional decay channel that can be ``fast" enough to be in contradiction with, among others, 
gamma-ray bounds: this implies that our model needs to assume a ``hadrophobic" structure \cite{Donini:1997yu,Hsieh:2010zr}, in order to stabilize this decay channel. 
The IceCube flux implies a severe suppression of the coupling to left-handed (LH) lepton and Higgs doublets, preventing $N_1$ production in the early Universe by means of this type of interactions. Instead, the additional $SU(2)_R$ gauge interactions in the LRSM can provide a viable way to produce $N_1$ without spoiling its stability. This is accomplished by means of a freeze-in mechanism \cite{Hall:2009bx}, tuned to reproduce the correct present-day DM abundance. In fact,
we found that the 
freeze-out production mechanism in the context of 
a LRSM does not lead to the correct relic abundance for PeV-scale DM, since it would require an extremely large entropy dilution. This mechanism has been instead successfully used for keV sterile neutrino DM~\cite{Adhikari:2016bei}, 
when sufficient entropy dilution can be achieved~\cite{Nemevsek:2012cd,bezrukov_lindner1}. See instead Ref.~\cite{Heeck:2015qra} 
for a discussion of TeV-scale DM in the context of the LRSM and Ref.~\cite{Merle:2013wta,Merle:2015oja} for the freeze-in mechanism in the 
context of a keV sterile neutrino DM. 

The next-to-lightest $N_2$ and the heaviest $N_3$ neutrinos in our model are responsible, in turn, for the generation of the matter-antimatter asymmetry of the Universe, via 
standard thermal leptogenesis. Specifically, we rely on the $N_2$-dominated scenario of leptogenesis \cite{geometry}, 
in which $N_2$'s dynamics is able to produce the correct final baryon asymmetry, measured by the baryon-to-photon ratio $\eta^\textup{CMB}_B=(6.1\pm0.1)\times 10^{-10}$ \cite{Planck}. In this case, the thermal production of $N_2$ relies on the Yukawa couplings to Higgs and LH doublets, which are not suppressed.

We therefore propose a LRSM in which the lightest heavy neutrino will play the role of the DM particle, while being also responsible, through its decays, 
for the IceCube signal. At the same time, the other two heavy neutrinos will generate the baryon asymmetry of the Universe via standard thermal leptogenesis. 

The paper is organized as follows: in Sec.~\ref{sec:model} we present the details of the LRSM model considered in the analysis, in Sec.~\ref{sec:N1} we focus on the DM particle of the model, the right-handed neutrino $N_1$, considering in particular the constraints from IceCube and 
the relic abundance. In Sec.~\ref{sec:asymmetry} we present the calculation for the baryon asymmetry and in Sec.~\ref{sec:conclusions} we draw our conclusions. 

\bigskip

\section{The model}\label{sec:model}
Considering the standard minimal LRSM \cite{lr_review1,lr_review2,senjanovic}, the RH leptons are fitted into $SU(2)_R$ doublets:
\begin{equation}
R_i=
\begin{pmatrix}
N_{Ri}\\
\ell_{Ri}
\end{pmatrix}.
\end{equation}
To ensure the right spontaneous symmetry breaking pattern and a Majorana mass term for the RH neutrinos, we consider a scalar field $\Delta_R$, triplet of $SU(2)_R$, which is responsible for the breaking to the Standard Model (SM) gauge group. The left-right symmetry implies the existence of an $SU(2)_L$ triplet $\Delta_L$. The electroweak symmetry breaking (EWSB) is then obtained by exploiting a bi-doublet scalar field $\Phi$. The Yukawa sector then reads:
\begin{equation}
\label{eq:lagrangian}
\mathcal{L}_Y=-Y^{(1)}_{ij}\overline{L}_i\Phi R_j-Y^{(2)}_{ij}\overline{L}_i\tilde{\Phi}R_j-Y^\Delta_{ij}\!\left(L_i^TCi\tau_2\Delta_LL_j+R_i^TCi\tau_2\Delta_RR_j\right)+\mbox{h.c},
\end{equation}
where 
\begin{equation}
\Phi=
\begin{pmatrix}
\phi_1^0 & \phi_1^+\\
\phi_2^- & \phi_2^0
\end{pmatrix},
\qquad
\Delta_{L,R}=
\begin{pmatrix}
\frac{1}{\sqrt{2}}\delta^+ & \delta^{++}\\
\delta^0 & -\frac{1}{\sqrt{2}}\delta^+
\end{pmatrix}_{L,R},
\end{equation}
and $\tilde{\Phi}\equiv\tau_2\Phi^*\tau_2$ ($\tau_2$ being the second Pauli matrix).

In order to avoid unwanted low-energy effects due to the $SU(2)_R$ gauge interactions, 
the $SU(2)_R\otimes SU(2)_L\otimes U(1)_{\tilde{Y}}\rightarrow SU(2)_L\otimes U(1)_Y$ breaking must take place at very high energies. Therefore, the triplet $\Delta_R$ acquires a vacuum expectation value (VEV) $\langle\Delta_R\rangle\equiv v_R$ which is suitably large. This implies a Majorana mass matrix for the RH neutrinos given by:
\begin{equation}
M_{ij}=2Y^\Delta_{ij}\,v_R,
\end{equation}
which can then be diagonalised to $D_M=\mbox{diag}(M_1,\,M_2,\,M_3)$.
We shall assume $M_{\Delta_R}\gg M_i$.
The VEV $v_R$ also sets the mass of the $SU(2)_R$ gauge bosons, since $m_{Z_R},\,m_{W_R}\propto v_R$.

Labelling the LH lepton doublet with the flavour index $\alpha=e,\mu,\tau$, the Yukawa couplings of the RH neutrinos with the SM Higgs doublet $H$ are given by:
\begin{equation}
\label{eq:RH_coupling}
\mathcal{L}_{Y^\nu}=-Y^\nu_{\alpha i}\,\overline{L}_\alpha H N_{Ri}+\mbox{h.c,}
\end{equation}
with
\begin{equation}
Y^\nu\equiv\frac{Y^{(1)}v_1+Y^{(2)}v_2}{\sqrt{v_1^2+v_2^2}},
\end{equation}
where $\langle\phi_i^0\rangle=v_i$. The SM Higgs VEV is obtained as: $\langle H\rangle=\sqrt{v_1^2+v_2^2}\equiv v\simeq174\,\mbox{GeV}$.
Finally, we may have that also $\Delta_L$ gets a VEV $v_L\ll v$. In general, we shall also assume $M_{\Delta_L}\gg M_i$.

With this symmetry-breaking pattern, the light neutrino masses are generally given by the combination of a type-I and a type-II seesaw terms. We will assume that the type-II seesaw contribution to the light neutrino masses is negligible, or even vanishing if $v_L=0$, so that the light neutrino mass matrix is given by:
\begin{equation}
m_\nu=-Y^\nu D_M^{-1}{Y^\nu}^T\,v^2.
\end{equation}
The light neutrino masses are then obtained through the PMNS mixing matrix $U$ as:
\begin{equation}
D_m=-U^\dagger\,m_\nu\,U^*,
\end{equation}
where $D_m=\mbox{diag}(m_1,\,m_2,\,m_3)$.

The most relevant interaction of the RH neutrinos are then given by the Yukawa coupling to Higgs and lepton doublets, in Eq.~\eref{eq:RH_coupling}, and by the $SU(2)_R$ gauge interactions. We will comment in subsection~\ref{subs:relic_abundance} on the possible interactions of the RH neutrinos with the RH triplet $\Delta_R$. 
If we assume the standard inflationary picture of the early Universe, charged leptons and LH neutrinos are part of the thermal bath, hence we can assume that $SU(2)_R$ interactions are able produce the RH neutrinos. In the absence of $SU(2)_R$ interactions, the production of the RH neutrinos is possible only through the Yukawa interactions, which become effective at temperatures around the RH neutrino mass.

\bigskip
\section{$N_1$ as the dark matter particle}\label{sec:N1}
\subsection{Constraints from IceCube}
In our model, we assume that the signal detected by IceCube is originated from DM decays. Given the mass pattern, the suitable DM candidate is the lightest heavy neutrino. Therefore, $N_1$ will be bound to constitute the whole DM content of the Universe and at the same time produce the IceCube signal.

Let us first analyse the constraints obtained from the IceCube data. $N_1$ decays only through the Yukawa couplings of Eq.~\eref{eq:RH_coupling}. 
The decay channels are:
\begin{equation}
N_1\longrightarrow l_\alpha^\mp W^\pm;\qquad N_1\longrightarrow\nu_\alpha Z,\bar{\nu}_\alpha Z;\qquad N_1\longrightarrow\nu_\alpha h,\bar{\nu}_\alpha h.
\label{eq:nudecay}
\end{equation}
For $M_1\gg m_Z,\,m_h$, we have monochromatic neutrinos with energy $E_\nu\simeq M_1/2$. This will cause a sharp peak and a cutoff in the neutrino energy spectrum, 
while neutrino cascades will provide a soft tail in the spectrum. From the highest event detected by IceCube \cite{icecube2,Aartsen:2015zva} 
we directly obtain the DM mass: $M_1\simeq4\,\mbox{PeV}$.

It is then possible to estimate the neutrino flux on Earth from the decay of $N_1$, assuming it constitutes the total amount of DM in the Universe. 
By comparing the theoretical prediction to the flux observed by IceCube, once the mass $M_1$ is set, 
it is possible to fix the $N_1$'s lifetime $\tau_{N_1}$. From Ref.~\cite{kitano}, we derive:
\begin{equation}
\tau_{N_1}\simeq 10^{28}\,\mbox{s}.
\label{eq:time}
\end{equation}
Although these are just approximate determinations of the mass and lifetime of $N_1$, as inferred from the IceCube data, for the purposes of this paper they can be regarded as a sufficiently good estimation: slight changes in these values will not make any noticeable difference.

The total decay rate $\Gamma_{D1}=\tau_{N_1}^{-1}$ at tree level is given as a function of the Yukawa parameters $Y^\nu_{\alpha1}$ by the expression \cite{davidson_nardi_nir,hambye}:
\begin{equation}
\label{eq:decay}
\Gamma_{D1}=\frac{M_1}{8\pi}\sum_\alpha\abs{Y^\nu_{\alpha1}}^2.
\end{equation}
Eq. (\ref{eq:time}) implies a constraint on the Yukawa couplings $Y^\nu_{\alpha1}$\footnote{From this result, it is evident that this setup cannot realize strong thermal leptogenesis \cite{initial_conditions,stl_lower_bound}.}:
\begin{equation}
\label{eq:yukawa_cond}
\sum_\alpha\abs{Y^\nu_{\alpha1}}^2=\frac{8\pi}{M_1\,\tau_{N_1}}\ll1.
\end{equation}
Due to the seesaw relation, this constraint will be reflected onto the other Yukawa couplings and the light neutrinos spectrum as well. It is then convenient to introduce the complex orthogonal matrix $\Omega$ parameterisation \cite{casas_ibarra}:
\begin{equation}
Y^\nu=\frac{1}{v}\,UD_m^{1/2}\,\Omega\,D_M^{1/2},
\end{equation}
such that:
\begin{equation}
\label{eq:y_omega}
\sum_\alpha\abs{Y^\nu_{\alpha1}}^2=\frac{M_1}{v^2}\sum_i m_i\abs{\Omega_{i1}}^2.
\end{equation}
Hence we obtain:
\begin{equation}
\label{eq:mtilde_cond}
\sum_i m_i\abs{\Omega_{i1}}^2=\frac{8\pi v^2}{M_1^2\,\tau_{N_1}}\simeq10^{-52}\,\mbox{eV}.
\end{equation}
Given the light neutrinos mass spectrum with nonzero $m_2$ and $m_3$, it is clear that, in order to have a vanishing $\sum_i m_i \abs{\Omega_{i1}}^2$, we must necessarily have $m_1\simeq0$ and a complex orthogonal matrix of the form:
\begin{equation}
\label{eq:omega}
\Omega\simeq
\begin{pmatrix}
1 & \beta\sin\theta-\alpha\cos\theta & \beta\cos\theta+\alpha\sin\theta\\
\alpha & \cos\theta & -\sin\theta \\
-\beta & \sin\theta & \cos\theta
\end{pmatrix},
\end{equation}
with $\alpha,\,\beta,\,\theta$ complex and $\abs{\alpha}\!,\abs{\beta}\simeq0$. All the Yukawa couplings, and hence all the quantities related to the other heavy neutrinos, can then be derived from a complex orthogonal matrix with the form in Eq.~\eref{eq:omega} and a fully hierarchical light neutrino spectrum. We also notice that the requirement $m_1\simeq0$ (in particular $m_1\ll 10^{-4}\,\mbox{eV}$) implies that this setup cannot be realised within the so-called $SO(10)$-inspired leptogenesis models \cite{buchmuller_plumacher,buccella_falcone_tramontano,nezri_orloff,branco,akhmedov,dibari_riotto1,dibari_riotto2,so10_decrypting}. \\
This special form of the complex orthogonal matrix $\Omega$ was pointed out in Ref. \cite{Anisimov:2008gg}, where a model with one vanishing eigenvalue in the Yukawa matrix was there presented. Thereby, the production  of the decoupled heavy neutrinos was obtained through active-sterile neutrino oscillations. Alternatively, it can be achieved through inflaton decay~\cite{Anisimov:2008gg,kitano}. 

To summarize, by imposing the bound on $N_1$'s lifetime from the IceCube flux has important consequences on the general setup of the model, due to the seesaw relation.

\bigskip

\subsection{Constraints on the model from $W_R$-mediated decays of $N_1$}

In addition to the decay mode into neutrinos listed in Eq. (\ref{eq:nudecay}) that allow an interpretation of the Ice Cube events, in the simplest realization of the LRSM where both right-handed leptons and right-handed quarks are accommodated in $SU(2)_R$ doublets, 
$N_1$ possesses also an hadronic decay through the mediation of the $W_R$ gauge boson into right-handed charged leptons and quarks: $N_1\rightarrow l_Rq_R\bar{q}_R'$. The decay rate of this process is \cite{Dev:2014iva}:
\begin{equation}
\Gamma(N_1\rightarrow l_Rq_R\bar{q}_R')=\frac{3 g_R^4}{2^9\pi^3M_1^3}\int_0^{M_1^2}\!ds\frac{M_1^6-3M_1^2s^2+2s^3}{\left(s-M_{W_R}^2\right)^2+M_{W_R}^4\frac{g_R^4}{(4\pi)^2}}.
\end{equation}
Considering $M_1\ll M_{W_R}$ and the usual condition $M_{W_R}=g_R v_R$, the decay rate can be cast in this
form:
\begin{align}
\Gamma(N_1\rightarrow l_Rq_R\bar{q}_R') \simeq \frac{3\,M_1^5}{2^{10}\pi^3\,v_R^4}
\end{align}
where we have also considered $g_R \ll 4\pi$.

This decay rate implies a lifetime for $N_1$ larger than the age of the Universe for $v_R > 5 \times 10^{17}$ GeV. 
This would reflect into an additional bound on the results we will present in the next sections, nevertheless leaving the standard LRSM viable. However,
extrapolations to the PeV mass range of antiproton bounds on heavy-DM decays \cite{Garny:2012vt} and bounds from 
gamma-rays \cite{Murase:2012xs,Esmaili:2015xpa} set much stronger constraints on this decay channel. 
Even though extrapolations of the knowledge of hadronization processes at such large energies and astrophysical uncertainties are likely present, 
nevertheless the lifetime associated to this decay channel is plausibly larger than $10^{26}\, {\rm s} -10^{27}\, {\rm s}$. This pushes $v_R$ in a transplanckian 
regime, complemented by a suitable requirement $g_R< g_L$, in order to keep at least the mass of $W_R$ below the Planck scale. This solution makes the standard LRSM quite contrived.

\begin{table}[t]
\center
\begin{tabular}{|c|c|c|c|c|}
 \hline
  ~&  $SU(3)_C$ & $SU(2)_L$ & $SU(2)_R$ & $U(1)_{\tilde{Y}}$\\
 \hline
 $L $ & {\bfseries1} & {\bfseries 2} & {\bfseries 1} & $-1/2$\\
  $Q$ & {\bfseries3} & {\bfseries 2} & {\bfseries 1} & $1/6$\\
 \hline
 $R $ & {\bfseries1} & {\bfseries 1} & {\bfseries 2} & $-1/2$\\
 $u$ & {\bfseries3} & {\bfseries 1} & {\bfseries 1} & $2/3$\\
  $d$ & {\bfseries3} & {\bfseries 1} & {\bfseries 1} & $-1/3$\\
 \hline 
\end{tabular}
\caption{Multiplet assignment for the hadrophobic LRSM. $L$ and $R$ are fermionic doublets, $Q$ is a quark doublet and $u$ and $d$ stand for the up quark and down quark singlets.}
\label{tab:HP}
\end{table}

Instead, an ``hadrophobic" LR choice of the representation for the right-handed quarks prevents the decay of $N_1$ through $W_R$. 
This model accommodates the right-handed quarks into singlets of $SU(2)_R$, with a suitable choice of their $\tilde{Y}$ quantum number in order to satisfy the condition 
$Q = T_{3L} + T_{3R} + \tilde{Y}$. The assignments are summarised in Table \ref{tab:HP}.
This model has been considered in the literature in the past, even though for different purposes, see e.g. Refs.\cite{Donini:1997yu,Hsieh:2010zr}. 
A caveat is that the model should be then embedded in a more complete and final theory to cure the problem of anomalies that are present in the hadrophobic LR model. An example was presented in Ref.~\cite{Das:2015ysz} in the case of the ``leptophobic"LR model. 
Note that in the hadrophobic LR model, we need also a doublet to give mass to the quarks in the singlet representation, that cannot couple to the bidoublet. This however does not change the results of our analysis.

\subsection{Relic abundance}
\label{subs:relic_abundance}
As mentioned before, we are requiring $N_1$ to be the DM particle. We must therefore be able to produce the correct abundance. More specifically, given the current values of $\Omega_{DM}$, of the critical density $\rho_c$ and entropy $s_0$, 
the current DM abundance $Y_{DM}^0\equiv n_{DM}/s_0$ is:
\begin{equation}
\label{eq:Yfin}
Y_{DM}^0=\frac{\Omega_{DM}\,\rho_c}{M_1s_0}\simeq3.82\times10^{-10}\left(\frac{\mbox{GeV}}{M_1}\right)\simeq9.5\times10^{-17},
\end{equation}
for $M_1=4$~PeV.

In general, $N_1$ can be produced from the thermal bath through its interactions given by the $SU(2)_R$ gauge bosons and the Yukawa couplings in Eq.~\eref{eq:RH_coupling}. We can safely neglect the contribution due to the coupling with $\Delta_R$. Indeed, assuming a very high scale ${v_R\gtrsim10^{14}\,\mbox{GeV}}$, while having $M_1\simeq4\,\mbox{PeV}$, implies that the Yukawa couplings $Y^\Delta$ for $N_1$ are extremely small and the interactions with $\Delta_R$ are strongly suppressed. Also the mixing of $N_1$ with the other heavy neutrinos gives a negligible contribution~\cite{Anisimov:2008gg}, 
as well as the decays $N_{2,3}\longrightarrow N_1l_Rl_R'$, mediated by $W_R$.

The Yukawa interactions with Higgs and lepton doublets give a decay rate at temperature $T$~\cite{pedestrians}:
\begin{equation}
\Gamma_{D1}(T)=\Gamma_{D1}\frac{\mathcal{K}_1(T)}{\mathcal{K}_2(T)},
\end{equation}
where $\mathcal{K}_i(T)$ are the modified Bessel functions of index $i$.

At the same time, the RH neutrinos are subject to the $SU(2)_R$ gauge interactions, whose scattering rate is given by:
\begin{equation}
\label{eq:gamma_S}
\Gamma_S(T)\equiv n^\textup{eq}_{N_1}(T)\langle\sigma\abs{v}\rangle(T),
\end{equation}
where $n_{N_1}^\textup{eq}$ is the equilibrium number density of $N_1$ and $\langle\sigma\abs{v}\rangle$ is the thermally averaged cross section times velocity.
The latter can be estimated via the usual neutrino scattering cross section, as \cite{bezrukov_lindner1}:
\begin{equation}
\langle\sigma \abs{v}\rangle(T)\simeq G_F^2T^2\left(\frac{m_W}{m_{W_R}}\right)^4\sim G_F^2T^2\left(\frac{m_W}{v_R}\right)^4,
\end{equation}
where $m_W$ is the $W$ boson mass and $G_F$ the Fermi constant.
Both $\Gamma_{D1}$ and $\Gamma_S$ enter the Boltzmann equations that describe the production of $N_1$. To this aim we can consider the variable $z\equiv M_1/T$ and the number density $N_{N_1}$ of RH neutrinos $N_1$, computed in a comoving volume containing one heavy neutrino $N_1$ in ultra-relativistic equilibrium. This can be easily related to the abundance $Y_{N_1}$. From the definition ${N_{N_1}\equiv n_{N_1}(T)/n_{N_1}^\textup{eq}(T\gg M_i)}$, we have $N_{N_1}=4/3\,n_{N_1}(T)/n_\gamma^\textup{eq}(T)$, where $n_\gamma^\textup{eq}$ is the equilibrium number density of photons. This is strictly connected to the entropy $s(T)=\pi^4 g_*^s(T)\,n_\gamma^\textup{eq}(T)/45\zeta(3)$, such that: 
\begin{equation}
N_{N_1}=\frac{4}{135}\frac{\pi^4 g_*^s(T)}{\zeta(3)}\frac{n_{N_1}(T)}{s(T)}\simeq 2.40g_*^s(T)\,Y_{N_1}(T).
\end{equation}
In the temperature range of our interest $g_*^s=g_*=112$, considering the three heavy neutrinos as relativistic.

Using $z$ and $N_{N_1}$ we can write \cite{pedestrians}:
\begin{equation}
\label{eq:boltzmann_production}
\frac{d N_{N_1}}{dz}=-(D_1(z)+S(z))\left[N_{N_1}(z)-N^\textup{eq}_{N_1}(z)\right],
\end{equation}
where $D(z)$ accounts for the decay/inverse-decay processes:
\begin{equation}
\label{eq:D}
D_1(z)\equiv\frac{\Gamma_{D1}(z)}{H(z)z},
\end{equation}
and $S(z)$ for the scattering:
\begin{equation}
\label{eq:S}
S(z)\equiv\frac{\Gamma_S(z)}{H(z)z}.
\end{equation}
Given Eqs.~\eref{eq:decay} and \eref{eq:yukawa_cond}, $N_1$'s decay rate is strongly suppressed, therefore we can just consider $S(z)$ in the 
equation for the evolution for $N_{N_1}$. 
Assuming vanishing initial abundance for $N_1$ at the end of inflation, i.e. $N_{N_1}(z_{RH})=0$ where $z_{RH}=M_1/T_{RH}$ corresponds to the 
reheating temperature, we can write the Boltzmann equation for $N_1$ as:
\begin{equation}
\label{eq:N1eq}
\frac{d N_{N_1}}{dz}=S(z)N_{N_1}^\textup{eq}(z),
\end{equation}
as long as $z<z_{eq}$, calling $z_{eq}$ the moment at which $N_1$ reaches the equilibrium distribution. Following Eq.~\eref{eq:S}, and adopting the expression of the Hubble rate in the radiation-dominated epoch \cite{kolb_turner}:
\begin{equation}
\label{eq:hubble}
H(z)=1.66g_*^{1/2}\frac{M_1^2}{M_{Pl}\,z^2},
\end{equation}
with $M_{Pl}$ being the Planck mass, we can easily find a solution:
\begin{equation}
N_{N_1}(z<z_{eq})=\frac{1}{4}\frac{\zeta(3)G_F^2\,M_{Pl}M_1^3}{1.66\pi^2\sqrt{g_*}}\left(\frac{m_W}{v_R}\right)^4\left(\frac{1}{z_{RH}^3}-\frac{1}{z^3}\right),
\end{equation}
where we used the expression of $S(z)$:
\begin{equation}
\label{eq:S(z)}
S(z)=\frac{3}{2}\frac{\zeta(3)G_F^2\,M_{Pl}M_1^3}{1.66\pi^2\sqrt{g_*}}\left(\frac{m_W}{v_R}\right)^4z^{-4}.
\end{equation}

\begin{figure}[!t]
\begin{tabular}{lr}
\hspace{-2cm}
\includegraphics[width=8.5cm,height=8cm]{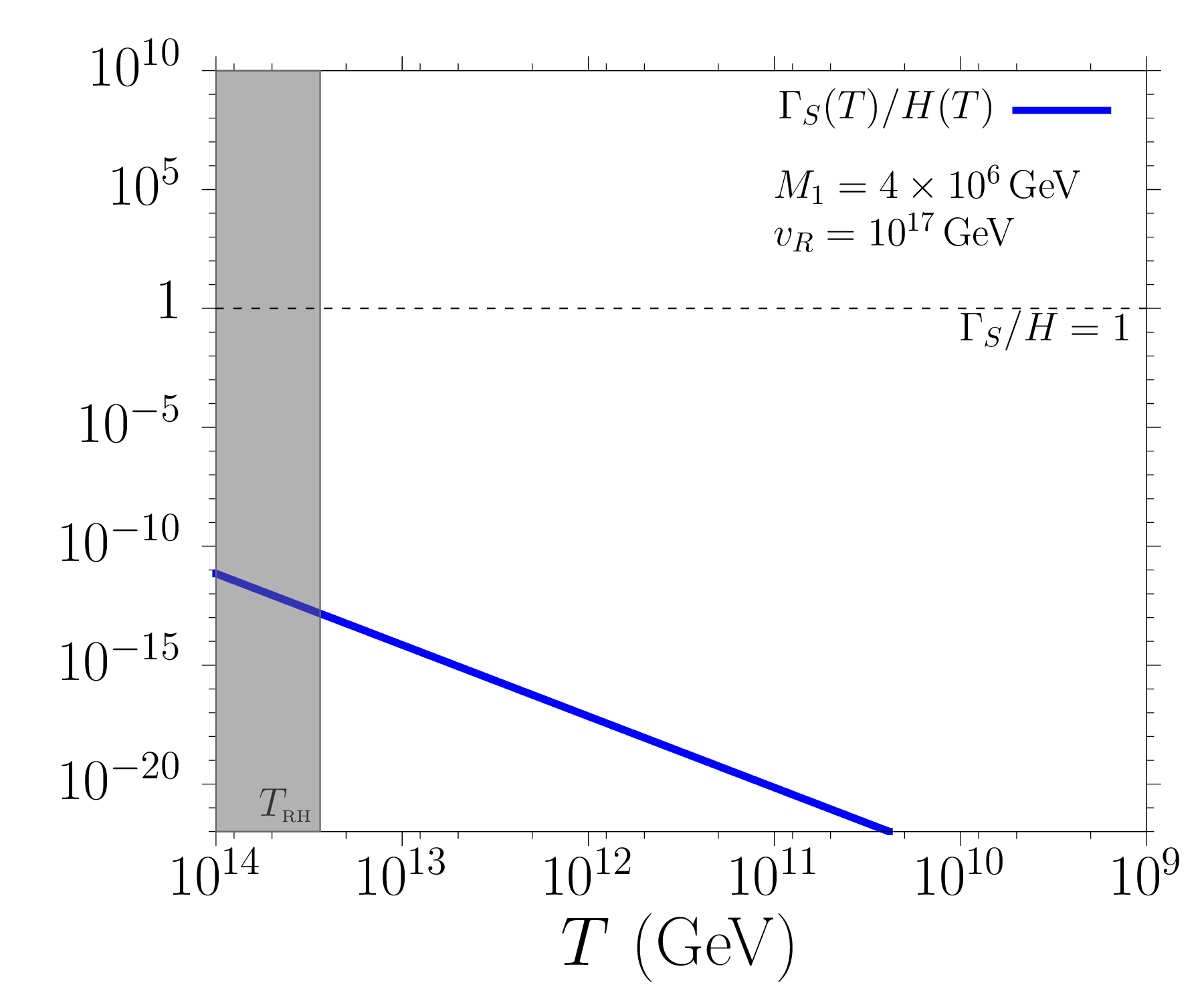} &
\includegraphics[width=8.5cm,height=8cm]{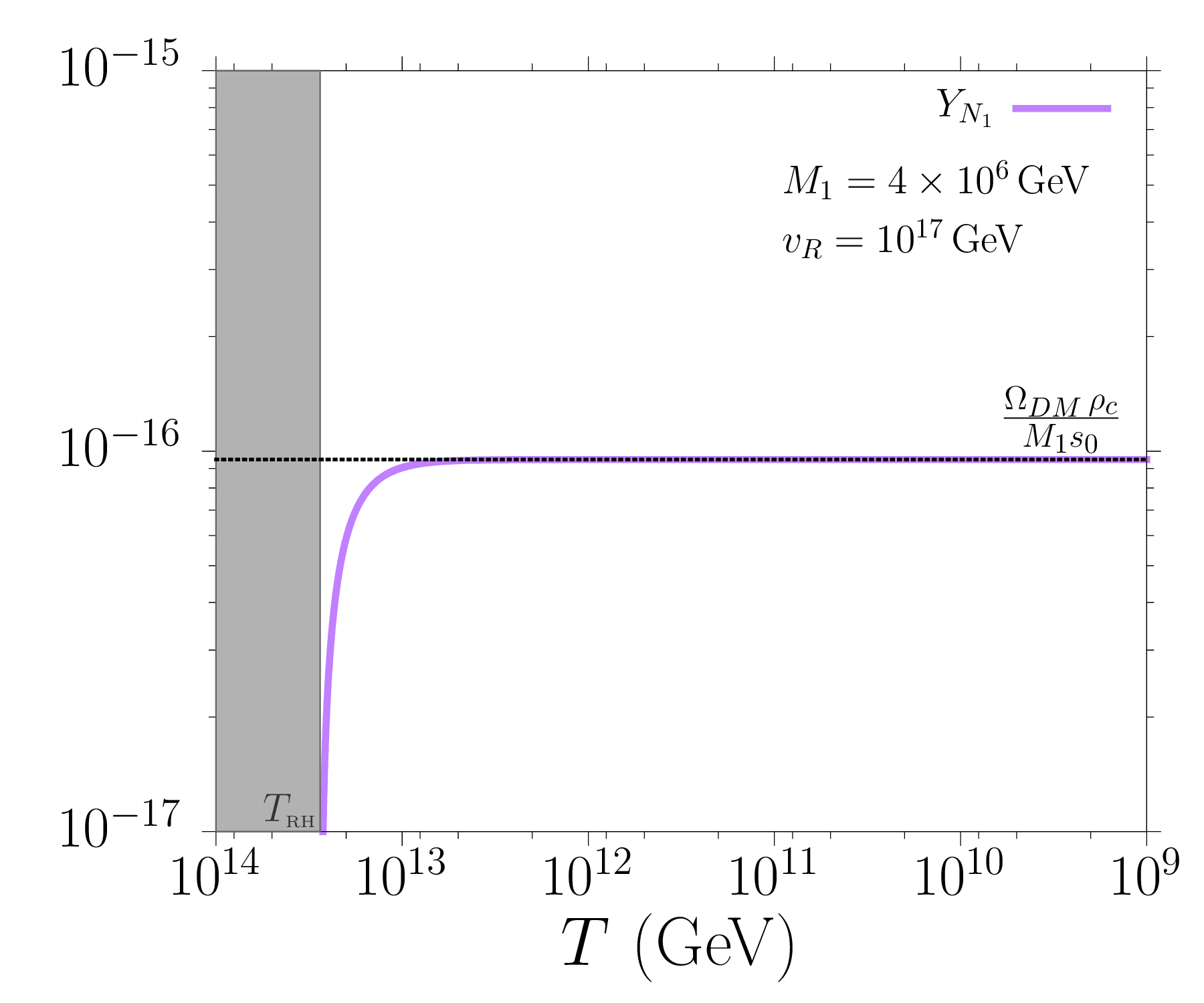} 
\end{tabular}
\caption{Left panel: Evolution in temperature of the ratio $\Gamma_S/H$ between scattering rate, Eq.~\eref{eq:gamma_S}, 
and the Hubble parameter, Eq.~\eref{eq:hubble}. Right panel: Evolution of $N_1$ abundance as a function of temperature.
We consider $M_1=4\times10^6\,\mbox{GeV}$, 
$T_{RH}=2.7\times10^{13}\,\mbox{GeV}$ and $v_R=10^{17}\,\mbox{GeV}$.
}
\label{fig:N1}
\end{figure}

\begin{figure}[!t]
\centering
\includegraphics[width=8.5cm,height=8cm]{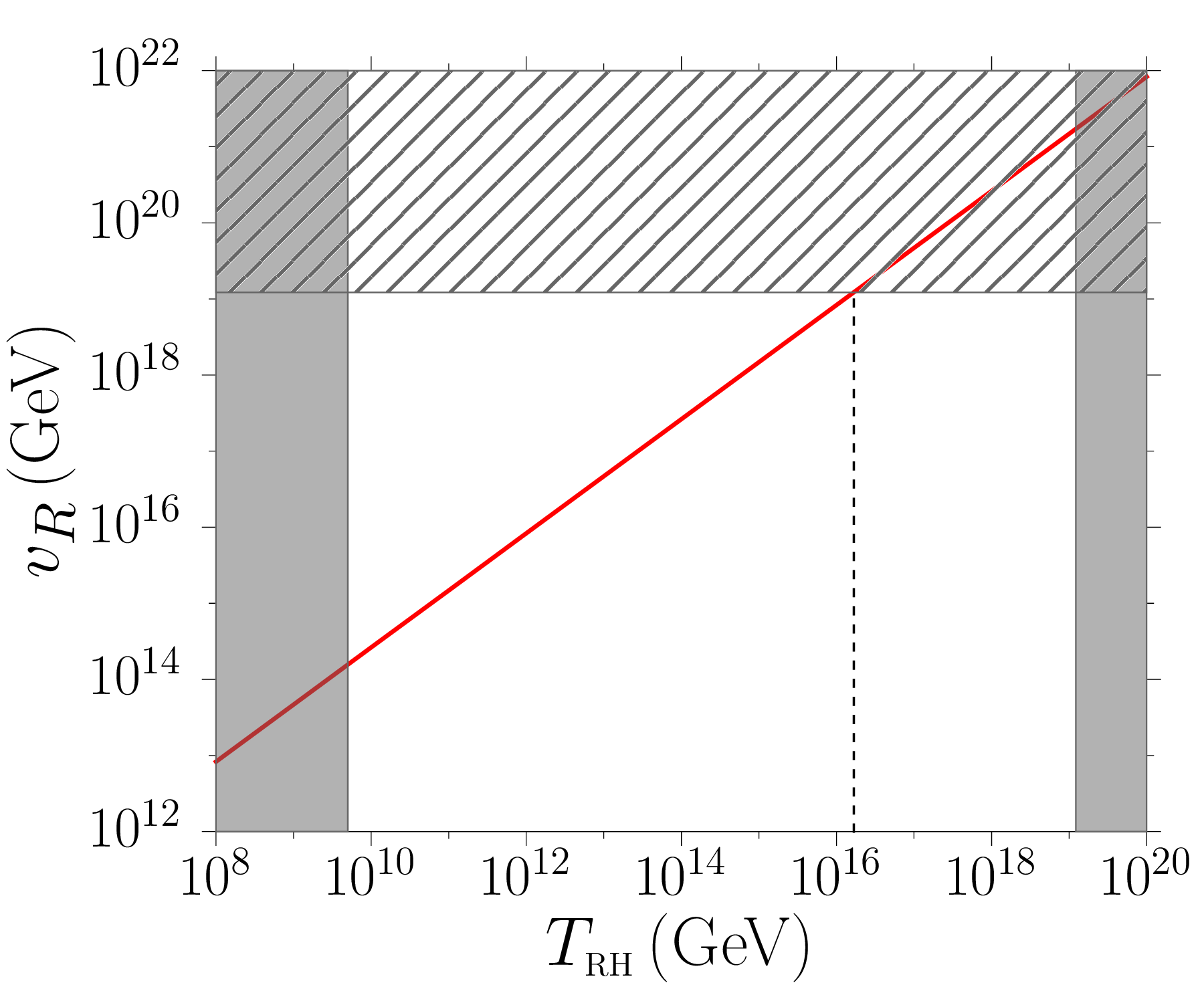}
\caption{Relation between $v_R$ and $T_{RH}$ that give the correct final abundance $Y_{N_1}^0=Y_{DM}^0$. The  vertical shadowed region on the right  
corresponds to temperatures above the Planck scale. 
The upper hatched region is cut out if transplanckian values of $v_R$ are excluded. 
The vertical shaded region on the left shows the lower bound given by leptogenesis, see the discussion 
in Sec.~\ref{sec:expressions_asym} and in particular Eq.~\eref{eq:TRH_bound}}. 
\label{fig:TRHvR}
\end{figure}

Given the expression of $S(z)$, it is clear that the scattering rate quickly decreases with the temperature, therefore we may expect an intermediate value $\bar{z}$, such that 
$z_{RH}<\bar{z}<z_{eq}$ at which $S(z)$ becomes negligible and the abundance of $N_1$ {\itshape freezes-in}, without reaching its 
equilibrium value. In this way, the current number density of $N_1$ is given by:
\begin{equation}
N^0_{N_1}\equiv N_{N_1}(\bar{z})\simeq\frac{1}{4}\frac{\zeta(3)G_F^2\,M_{Pl}M_1^3}{1.66\pi^2\sqrt{g_*}}\left(\frac{m_W}{v_R}\right)^4\frac{1}{z_{RH}^3}.
\end{equation}
In the left panel of Fig.~\ref{fig:N1} we show the ratio $\Gamma_S(z)/H(z)$, that measures 
the efficiency of the scattering reaction. As can be seen from the figure, this ratio is well below one even at the reheating temperature $T_{RH}$. 
The abundance of $N_1$ as a function of temperature $T$ 
is shown in the right panel of Fig.~\ref{fig:N1}. 

This model allows for the production of a relatively small abundance of RH neutrinos, in particular of $N_1$, exploiting the freeze-in mechanism. 
The scattering processes mediated by $SU(2)_R$ gauge bosons never become efficient after inflation, since their freeze-out temperature is always higher 
than $T_{RH}$. Thus, after reheating, the abundance of $N_1$ can only reach a 
low final value, which is then preserved due to the absence of any other interaction.

From Eq.~\eref{eq:Yfin}, we can obtain a relation between $v_R$ and the reheating temperature $T_{RH}$ such that the final abundance of $N_1$ is equal to the current DM abundance. We have:
\begin{equation}
Y^0_{N_1}\simeq\frac{1}{4\times2.40g_*}\frac{\zeta(3)G_F^2\,M_{Pl}}{1.66\pi^2\sqrt{g_*}}\left(\frac{m_W}{v_R}\right)^4T_{RH}^3,
\end{equation}
and imposing the value in Eq.~\eref{eq:Yfin} we get:
\begin{equation}
\label{eq:TRHvR}
v_R=\left[\frac{1}{4\times2.40g_*}\frac{\zeta(3)G_F^2 M_{Pl}}{1.66\pi^2g_*^{1/2}\,Y_{DM}^0}\right]^\frac{1}{4} m_W\,T^{3/4}_{RH}\simeq8.3\times10^6\,T_{RH}^{3/4},
\end{equation}
This relation is a prediction of our model, and it is shown in Fig.~\ref{fig:TRHvR}.

\bigskip
\section{The baryon asymmetry}\label{sec:asymmetry}

In our model the other two heavy neutrinos $N_2$ and $N_3$ are responsible for the production of the correct amount of baryon asymmetry of the Universe.
Since we assume that $\Delta_L$ does not contribute to the light neutrino masses and that $M_{\Delta_L}\gg M_i$, our model reduces to ordinary RH neutrino leptogenesis, 
without any effect from the triplet \cite{hambye}. For a discussion of the situations in which the scalar triplet could 
affect leptogenesis we refer to Refs.~\cite{hambye,Arina:2011cu}.

Given the constraint on $N_1$'s Yukawa couplings in Eq.~\eref{eq:yukawa_cond}, that completely decouples the lightest neutrino $N_1$, and the fact that $M_1\ll10^9\,\mbox{GeV}$ \cite{davidson_ibarra}, the asymmetry must be produced by $N_2$. Since $N_1$ does not play any role here, we can already expect on $M_2$ the same lower bound that applies on $M_1$ in $N_1$-dominated leptogenesis, i.e. $M_2\gtrsim 5\times10^8\,\mbox{GeV}$ \cite{davidson_ibarra,new_bounds}. For definiteness,
we will always consider hierarchical leptogenesis, obtained by imposing $M_3\geq3M_2$. Assuming $M_2\gtrsim5\times10^8\,\mbox{GeV}$, the asymmetry can be produced in two different regimes according to the mass of $N_2$ \cite{density,flavour_effects,fuller}: for $5\times 10^8\,\mbox{GeV}\lesssim M_2\lesssim 5\times 10^{11}\,\mbox{GeV}$ leptogenesis will take place in a two fully-flavoured regime, while for $M_2\gtrsim 5\times 10^{11}\,\mbox{GeV}$ leptogenesis will be unflavoured. 

\bigskip
\subsection{Asymmetry production}

\underline{{\it Case I:~~~$5\times10^8\,{\rm GeV}\lesssim M_2\lesssim 5\times10^{11}\,{\rm GeV}$}} \\
At temperatures $T\lesssim5\times 10^{11}\,\mbox{GeV}$ the charged $\tau$-Yukawa interactions are in equilibrium and more efficient than the LH-RH neutrino interactions \cite{new_bounds,zeno_effect}. 
Therefore, the relevant quantities will be the asymmetries $\Delta_\tau\equiv B/3-L_\tau$ and $\Delta_{\tau_2^\bot}\equiv B/3-L_{\tau_2^\bot}$, where $\tau_2^\bot$ defines the flavour component produced by $N_2$, orthogonal to the $\tau$ flavour direction.
Neglecting\footnote{The impacts of the corrections we neglect have been estimated not to be larger that about $20\%$~{\cite{new_bounds,quantum_riotto2}}.} flavour coupling \cite{abada2,nardi_nir_racker_roulet2,barbieri_creminelli_strumia_tetradis,davidson_nardi_nir,flavour_effects,spectator_processes,fuller}, scattering terms~\cite{Fong:2010bh,pedestrians,new_bounds}, thermal effects~\cite{thermal_effects} and quantum corrections~\cite{quantum_buchmuller,quantum_riotto1,quantum_riotto2}, the $B-L$ asymmetry produced by $N_2$, $N_{B-L}^\textup{lep,2}$, is given by the sum of the produced $\Delta_\tau$ and $\Delta_{\tau_2^\bot}$ asymmetries as \cite{fuller}
\begin{align}
N_{B-L}^\textup{lep,2}&=N_{\Delta_\tau}^\textup{lep,2}+N_{\Delta_{\tau_2^\bot}}^\textup{lep,2}\nonumber\\
\label{eq:asymm_case1}
&\simeq\varepsilon_{2\tau}\kappa_f(K_{2\tau})+\varepsilon_{2\tau^\bot}\kappa_f(K_{2\tau_2^\bot}).
\end{align}
Here, $\varepsilon_{2\tau^\bot}=\varepsilon_{2e}+\varepsilon_{2\mu}$, where $\varepsilon_{2\alpha}$ ($\alpha=e,\,\mu,\,\tau$) are the $N_2$ $C\!P$ asymmetries in flavour $\alpha$ defined as:
\begin{equation}
\varepsilon_{2\alpha}\equiv-\frac{\Gamma_{2\alpha}-\overline{\Gamma}_{2\alpha}}{\Gamma_2+\overline{\Gamma}_2},
\end{equation}
where $\Gamma_{2\alpha}$ and $\overline{\Gamma}_{2\alpha}$ are the rates of the decays $N_2\rightarrow l_\alpha H$ and $N_2\rightarrow\overline{l}_\alpha H$ respectively and $\Gamma_2=\sum_\alpha \Gamma_{2\alpha}$, $\overline{\Gamma}_2=\sum_\alpha \overline{\Gamma}_{2\alpha}$.
We also introduced the flavoured decay parameters:
\begin{align}
K_{i\alpha}&\equiv\frac{\Gamma_{i\alpha}+\overline{\Gamma}_{i\alpha}}{H(T=M_i)}\nonumber\\
\label{eq:ki}
&=\frac{\abs{Y^\nu_{\alpha i}}^2v^2}{m_*M_i}=\frac{1}{m_*}\abs{\sum_j \sqrt{m_j}\,U_{\alpha j}\Omega_{ji}},
\end{align}
where $m_*$ is the equilibrium neutrino mass \cite{pedestrians}, defined as
\begin{equation}
m_*=\frac{8\pi v^2}{H(T=M_i)}=\frac{16\pi^{5/2}g_*^{1/2}}{3\sqrt{5}}\frac{v^2}{M_{Pl}}\simeq1.08\times10^{-3}\,\mbox{eV}.
\end{equation}

In Eq.~\eref{eq:asymm_case1} we have $K_{2\tau_2^\bot}=K_{2e}+K_{2\mu}$, while the $\kappa_f(K_{2\alpha})$ are the efficiency factors, whose 
expressions can be found in the literature \cite{pedestrians,vives,new_bounds}. \\

\noindent
\underline{{\it Case II:~~~$M_2>5\times10^{11}\,{\rm GeV}$ }}\\
In this case, the flavour interactions are not in equilibrium, therefore the coherence of the lepton and anti-lepton states produced by $N_2$ is not broken. 
Therefore, the evolution of the full $B-L$ asymmetry and $N_2$'s abundance is tracked by the Boltzmann equations, whose solution gives
\begin{equation}
\label{eq:asymm_case2}
N_{B-L}^\textup{lep,2}\simeq\varepsilon_2\kappa_f(K_2),
\end{equation}
where $K_2=\sum_\alpha K_{2\alpha}$ is the total $N_2$'s decay parameter.
\null\\

\noindent We now have an expression for the $B-L$ asymmetry produced by $N_2$'s dynamics, both for Case I and Case II. 
In order to obtain the final value of the asymmetry we still have to take into account the impact of the processes involving $N_1$.

\bigskip
\subsection{\label{sec:expressions_asym}Expressions for the final asymmetry}
Below $T\sim M_2$, the asymmetry stays constant. However, for temperatures $T\lesssim5\times10^8\,\mbox{GeV}$ the $\mu$ Yukawa interactions are in equilibrium.

In Case I, this implies that 
at temperatures $M_1<T'\lesssim5\times10^8\,\mbox{GeV}$ the asymmetry $N^\textup{lep,2}_{\Delta_{\tau_2^\bot}}$ gets projected onto the $e$ and $\mu$ flavour directions. Neglecting phantom terms \cite{fuller,density}, we obtain the asymmetries in $\Delta_\delta\equiv B/3-L_\delta$ ($\delta=e,\,\mu$), at temperature $T'$, simply by
\begin{equation}
\label{eq:projection1}
N_{\Delta_e}(T')=\frac{K_{2e}}{K_{2\tau_2^\bot}}N_{\Delta_{\tau_2^\bot}}^\textup{lep,2},\qquad N_{\Delta_\mu}(T')=\frac{K_{2\mu}}{K_{2\tau_2^\bot}}N_{\Delta_{\tau_2^\bot}}^\textup{lep,2}.
\end{equation}
The action of $N_1$ will then take place along the three flavour directions $e,\mu,\tau$. Considering that the asymmetry produced by $N_1$ can be safely neglected, neglecting again flavour coupling and taking as initial conditions $N_{\Delta_e}$, $N_{\Delta_\mu}$ in Eq.~\eref{eq:projection1} and $N_{\Delta_\tau}$ in Eq.~\eref{eq:asymm_case1}, the Boltzmann equations can be solved, for Case I, giving the total final asymmetry as the sum of the final asymmetries in $\Delta_\alpha$, $\alpha=e,\,\mu,\,\tau$ \cite{vives,new_bounds,dibari_riotto1,density,fuller}:
\begin{align}
\label{eq:n1_case1}
N_{B-L}^\textup{lep,f}&= \sum_{\alpha=e,\,\mu,\,\tau} N_{\Delta_\alpha}^\textup{lep,f}\nonumber\\
&\simeq\frac{K_{2e}}{K_{2\tau_2^\bot}}\varepsilon_{2\tau^\bot}\kappa_f(K_{2\tau_2^\bot})e^{-\frac{3\pi}{8}K_{1e}}+\frac{K_{2\mu}}{K_{2\tau_2^\bot}}\varepsilon_{2\tau^\bot}\kappa_f(K_{2\tau_2^\bot})e^{-\frac{3\pi}{8}K_{1\mu}}+\varepsilon_{2\tau}\kappa_f(K_{2\tau})e^{-\frac{3\pi}{8}K_{1\tau}}.
\end{align}
As for Case II, 
the asymmetry in Eq.~\eref{eq:asymm_case2} gets projected as:
\begin{equation}
\label{eq:projection2}
N_{\Delta_e}(T')=\frac{K_{2e}}{K_2}N_{B-L}^\textup{lep,2},\qquad N_{\Delta_\mu}(T')=\frac{K_{2\mu}}{K_2}N_{B-L}^\textup{lep,2},\qquad N_{\Delta_\tau}(T')=\frac{K_{2\tau}}{K_2}N_{B-L}^\textup{lep,2},
\end{equation}
and, similarly, the final asymmetry is given by
\begin{align}
\label{eq:n1_case2}
N_{B-L}^\textup{lep,f}&= \sum_{\alpha=e,\,\mu,\,\tau} N_{\Delta_\alpha}^\textup{lep,f}\nonumber\\
&\simeq \frac{K_{2e}}{K_2}\varepsilon_2\kappa_f(K_2)e^{-\frac{3\pi}{8}K_{1e}}+
\frac{K_{2\mu}}{K_2}\varepsilon_2\kappa_f(K_2)e^{-\frac{3\pi}{8}K_{1\mu}}+
\frac{K_{2\tau}}{K_2}\varepsilon_2\kappa_f(K_2)e^{-\frac{3\pi}{8}K_{1\tau}}.
\end{align}

As already pointed out, in our specific model, $N_1$'s Yukawa couplings are suppressed. Therefore its washout is negligible and the final asymmetry expressions can be simplified. Indeed, from Eqs.~\eref{eq:ki} and \eref{eq:yukawa_cond} we have $K_{1\alpha}\simeq0$.
For this reason, from Eq.~\eref{eq:n1_case1}, we obtain for Case I:
\begin{equation}
\label{eq:final_case1}
N_{B-L}^\textup{lep,f}\simeq\varepsilon_{2\tau^\bot}\kappa_f(K_{2\tau_2^\bot})+\varepsilon_{2\tau}\kappa_f(K_{2\tau}),
\end{equation}
while for Case II, from Eq.~\eref{eq:n1_case2}, we get:
\begin{equation}
\label{eq:final_case2}
N_{B-L}^\textup{lep,f}\simeq \varepsilon_2\kappa_f(K_2).
\end{equation}
We notice that, due to the negligible washout by $N_1$, the phantom terms would anyway cancel out, therefore these final expressions are not affected by  this correction.
\medskip

\begin{figure}[!t]
\begin{tabular}{lr}
\hspace{-2cm}
\includegraphics[width=8.5cm,height=8cm]{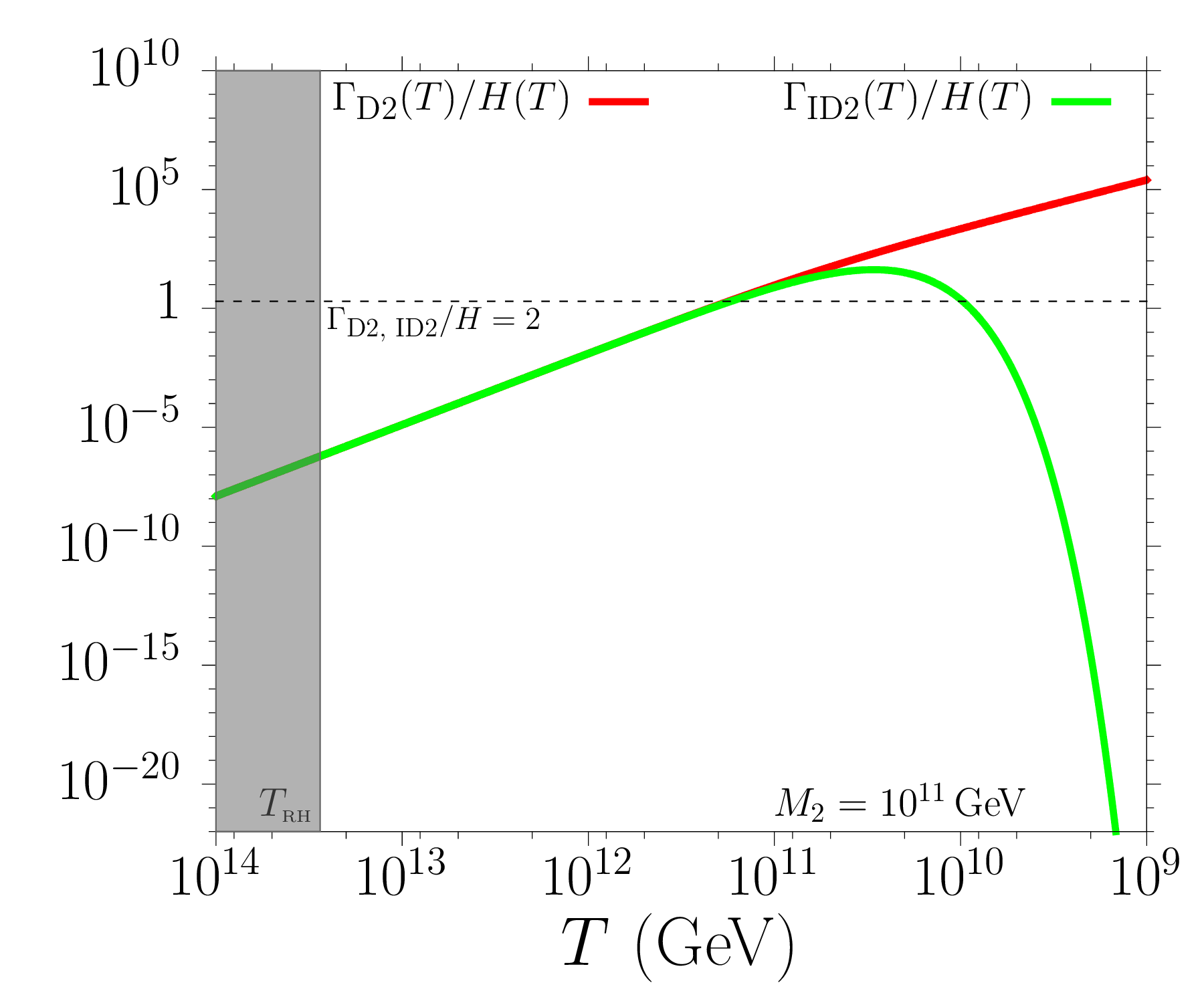} &
\includegraphics[width=8.5cm,height=8cm]{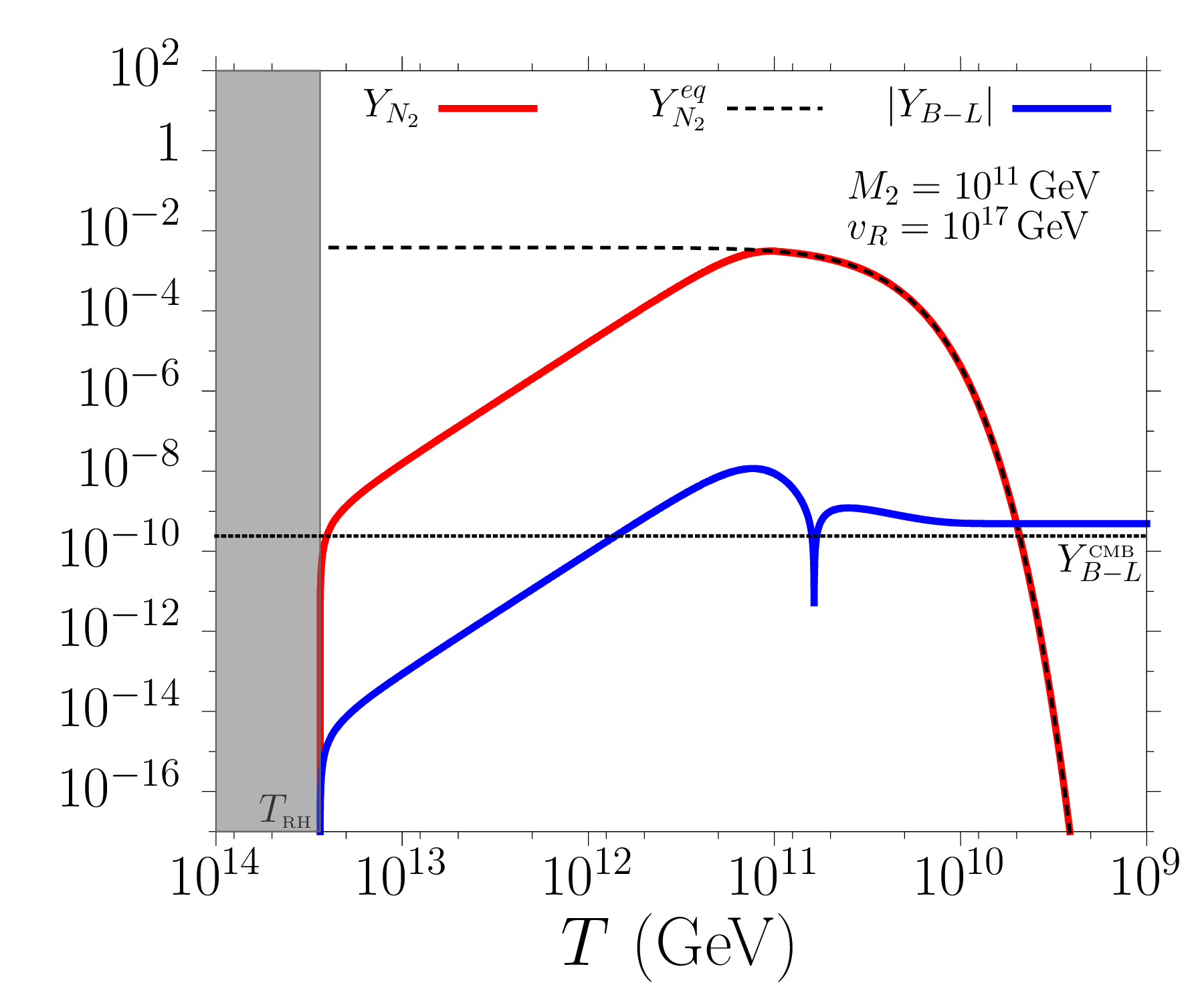}
\end{tabular}
\caption{
Left panel: Evolution in temperature of the ratio between $N_2$'s decay and inverse decay rates and the Hubble parameter. 
Right panel: Evolution of $N_2$ and $B-L$ asymmetry abundances as a function of temperature.
We consider $M_2=10^{11}\,\mbox{GeV}$, 
$T_{RH}=2.7\times10^{13}\,\mbox{GeV}$ and $v_R=10^{17}\,\mbox{GeV}$. The final $B-L$ asymmetry abundance is obtained as 
$Y^\textup{CMB}_{B-L}=\eta_B^\textup{CMB}/(2.40g_*\,0.96\times10^{-2})$, accounting for the sphaleron conversion rate and the dilution factor.}
\label{fig:N2_BL}
\end{figure}

\begin{figure}[!t]
\centering
\includegraphics[height=8cm,width=10cm]{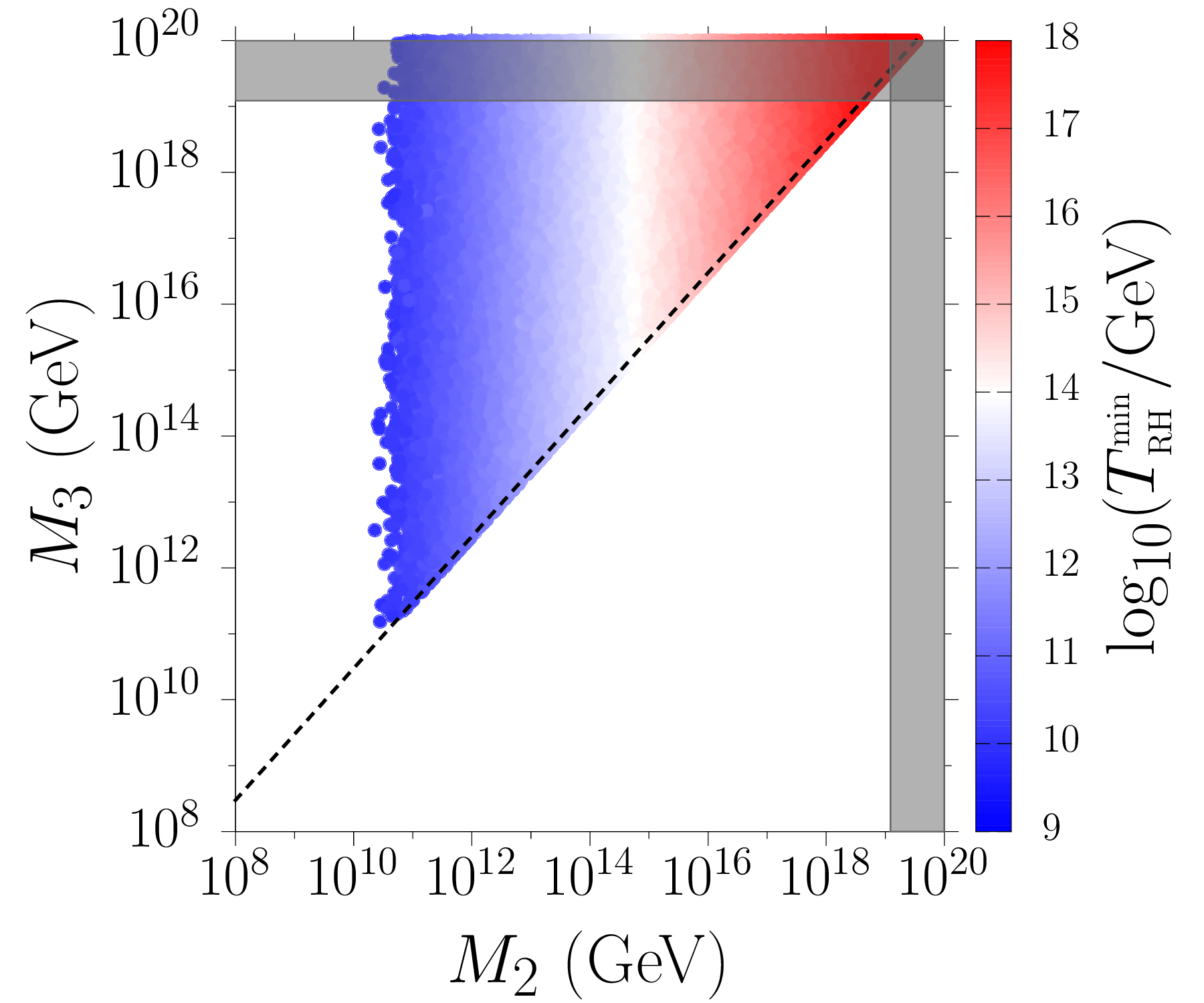}
\caption{Scatter plot of points in the plane $M_2$-$M_3$ that give successful leptogenesis. The colour gives the corresponding 
$\log_{10}(T_{RH}^\textup{min}/\mbox{GeV})$. The dashed line shows the hierarchical limit $M_3=3M_2$. The shaded regions exclude masses $M_i\geq M_{Pl}$.} 
\label{fig:M2M3}
\end{figure}

As an example, in Fig.~\ref{fig:N2_BL} we consider the case with $M_2=10^{11}\,\mbox{GeV}$ and a particular choice of the matrix $\Omega$ in Eq.~\eref{eq:omega}. 
In the left panel, we show the ratios $\Gamma_{D2}(z)/H(z)$ and $\Gamma_{ID2}(z)/H(z)$. 
These ratios reach a value equal to two at a temperatures between $10^{11}-10^{12}$~GeV. 
In the right panel, we show the evolution of $N_2$ and $B-L$ asymmetry abundances. 
As can be seen, the $N_2$'s abundance (red line) can reach and track the equilibrium 
distribution (dashed line), as in the strong-washout regime \cite{pedestrians,fuller}. 
The blue line marks the evolution of the asymmetry abundance and it is clear that the final value is above the experimental 
bound $Y_{B-L}^\textup{CMB}$. Therefore, we can conclude that it is possible, in our model, to obtain successful leptogenesis from $N_2$'s decays. 
Note that in the case shown in the figure, the asymmetry is produced in the two-flavour regime, i.e. Case I Eq.~\eref{eq:final_case1}. 

In Fig.~\ref{fig:M2M3} a general scan on $M_2$ and $M_3$, with $M_3\geq3M_2$ is performed and Case I and Case II (Eqs.~\eref{eq:final_case1} and \eref{eq:final_case2} respectively) are adopted according to the value of $M_2$. Here we assume Normal 
Ordering (NO) of the light neutrino masses $m_2^2=m_1^2+\msol^2$, $m_3^2=m_1^2+\matm^2$ with $\msol=0.0087\,\mbox{eV}$ 
and $\matm=0.0496\,\mbox{eV}$ \cite{maltoni}\footnote{See also Ref.~\cite{Gonzalez-Garcia:2015qrr} for a recent review on neutrino oscillation 
parameters.}. According to our previous discussion, we set $m_1=0$. We scan on the neutrino mixing angles uniformly extracting them in their experimental $3\sigma$ ranges as in \cite{maltoni}, while the Dirac and Majorana phases are extracted on their full variability range. We also scan on the complex angle $\theta$ in $\Omega$, while setting $\alpha=\beta=0$ for simplicity. Moreover, we require $\abs{\Omega_{ij}}^2\leq2$.

We assume here an instantaneous transition from the two fully-flavoured to the unflavoured regime, at $M_2=5\times10^{11}\,\mbox{GeV}$. A more accurate description should employ a full density matrix formalism \cite{density} to describe leptogenesis in the transition region, however only a small impact is expected. It is also possible to find a lower bound on $M_2$ as we expected from considerations similar to $N_1$-dominated leptogenesis. In this case the bound appears to be slightly higher: ${M_2\gtrsim 10^{10}\,\mbox{GeV}}$.

In Fig.~\ref{fig:M2M3}, the colours encode the minimal reheating temperature $T^\textup{min}_{RH}$ needed to produce the correct final asymmetry for each value of $(M_2,\,M_3)$. This is obtained as \cite{pedestrians}:
\begin{equation}
T^\textup{min}_{RH}\simeq\frac{M_2}{z_L(\overline{K}_2)-2e^{-3/\overline{K}_2}},
\end{equation}
where $\overline{K}_2=K_2$ in Case II, while $\overline{K}_2=K_{2\tau},K_{2\tau_2^\bot}$ in Case I, if the asymmetry in the $\tau$ or $\tau_2^\bot$ flavour dominates respectively.
From Fig.~\ref{fig:M2M3} it is possible to notice that the lowest values are obtained for $M_2$ around the lower bound. We obtained a lower bound entirely given by leptogenesis: 
\begin{equation}
\label{eq:TRH_bound}
T^\textup{min}_{RH}\gtrsim 7\times10^9\,\mbox{GeV}.
\end{equation}
Using this lower bound on $T^\textup{min}_{RH}$ and Eq.~\eref{eq:TRHvR}, we can find a range of allowed values of $v_R$, 
see also Fig.~\ref{fig:TRHvR}. We have $2\times10^{14}\,\mbox{GeV}\lesssim v_R\lesssim 2\times10^{21}\,\mbox{GeV}$, or the more restrictive 
case ${2\times10^{14}\,\mbox{GeV}\lesssim v_R\lesssim 10^{19}\,\mbox{GeV}}$ if we require $v_R<M_{Pl}$. We notice that the lower 
bound on $v_R$ agrees with our assumptions of a very high symmetry breaking scale.
\section{Conclusions}\label{sec:conclusions}

In this paper, we have considered a left-right symmetric model, where a DM particle is produced through a freeze-in process. 
The model is able to produce the correct final abundance of DM and baryon asymmetry, while at the same time the DM candidate is suitably heavy, 
$M_1=4\,\mbox{PeV}$, and long-lived, $\tau_{N_1}\simeq 10^{28}\,\mbox{s}$, to produce high-energy neutrinos consistent with the IceCube signal. 
It is interesting to point out that this can be realised only with vanishing absolute neutrino mass scale $m_1\simeq0$, which allows for a long 
lifetime $\tau_{N_1}$ through tiny values of $N_1$'s Yukawa couplings. 
As a consequence, this model predicts a power-law relation between the reheating temperature of the Universe, $T_\textup{RH}$, and the 
vacuum expectation value of the $SU(2)_R$ triplet. To obtain successful leptogenesis, a lower bound 
$T_\textup{RH}\gtrsim7\times10^9\,\mbox{GeV}$ should be satisfied. 

We notice that the LRSM has been recently considered also in the context of the diphoton excess and diboson~\cite{Berlin:2016hqw}, 
considering a weak-scale DM produced through a thermal freeze-out. 
Within our setup the diphoton and diboson excess cannot be explained, because it would require $v_R \sim 3-4$~TeV, and 
thus a $W^\prime$ boson of mass of the order of $1.8-2$~TeV, and a $Z^\prime$ boson mass 
of $3-4$~TeV.

In conclusion, the model presented in this paper successfully provides a consistent solution to both the DM problem and the generation of the
matter-antimatter asymmetry of the Universe through leptogenesis, while producing at the same time a viable interpretation of the highest energy IceCube neutrino events. 
The basis is a left-right symmetric model, where the right-handed neutrino fields are all involved and mutually necessary in the generation of the different mechanisms 
at hand (correct DM relic abundance, baryon asymmetry and IceCube neutrino flux). The results can therefore be accommodated 
naturally in the LRSM scheme.

\clearpage

\section*{\label{sec:ack}Acknowledgements} 
We thank Bhupal Dev for relevant discussions. 
MRF acknowledges financial support from the STAG Institute and is also deeply grateful to the Physics Department of the University of Torino for the kind hospitality. 
VN acknowledges support by Spanish MINECO through project FPA2012-31880, by Spanish MINECO (Centro de excelencia Severo Ochoa Program) under grant SEV-2012-0249. 
VN acknowledges financial support by the European Union through the ITN ELUSIVES H2020-MSCA-ITN-2015//674896 and the RISE INVISIBLESPLUS H2020-MSCA-RISE-2015//690575. 
This work is supported by the research grant {\it Theoretical Astroparticle Physic}s number 2012CPPYP7 under the program PRIN 2012 funded by the Ministero dell'Istruzione, Universit\`{a} e della Ricerca (MIUR) and 
by the research grant {\it TAsP} (Theoretical Astroparticle Physics) funded by the Istituto Nazionale di Fisica Nucleare (INFN).

\bibliographystyle{JHEP}


\end{document}